\documentclass{iopart}

\usepackage{iopams}
\usepackage{setstack}
\usepackage{color}
\usepackage{graphicx}

\usepackage{url}

\def\mbf#1{\ensuremath{\mathchoice{\mbox{\boldmath$\displaystyle#1$}}
{\mbox{\boldmath$\textstyle#1$}}
{\mbox{\boldmath$\scriptstyle#1$}}
{\mbox{\boldmath$\scriptscriptstyle#1$}}}}

\def\dcc{\texttt{LIGO-P070030-02-Z}}
\def\aei{\texttt{AEI-2007-016}}

\def\const{\mathrm{const}}

\def\O{\mathcal{O}}

\def\F{\mathcal{F}}
\def\Fbar{\overline{\F}}

\def\parm{\lambda}
\def\vparm{\mbf{\parm}}
\def\dvec{\mbf}
\def\Parm{\mathbb{P}}
\def\Temp{\mathbb{T}}
\def\Eucl{\mathbb{E}}
\def\Z{\mathbb{Z}}
\def\R{\mathbb{R}}
\def\unit{\mathbb{I}}
\def\tr{\mathrm{T}}

\def\sig{\mathrm{s}}
\def\max{\mathrm{max}}

\def\vol{\mathrm{vol}}
\def\WS{\mathrm{WS}}
\def\FP{\mathrm{FP}}

\def\wt{\widetilde}
\def\Mt{\wt{M}}
\def\lt{\wt{l}}
\def\Mh{\widehat{M}}
\def\lh{\widehat{l}}

\def\mis{m}

\def\Lat{\Lambda}

\begin{document}

\title[Efficient lattice covering of flat parameter
spaces]{Template-based searches for gravitational waves: efficient
  lattice covering of flat parameter spaces}  
\author{Reinhard Prix}
\address{Max-Planck-Institut f\"{u}r Gravitationsphysik
  (Albert-Einstein-Institut), Am M\"uhlenberg 1, 14476 Golm, Germany}
\ead{reinhard.prix@aei.mpg.de}
\begin{abstract}
  The construction of optimal template banks for matched-filtering
  searches is an example of the \emph{sphere covering problem}.
  For parameter spaces with constant-coefficient metrics a
  \mbox{(near-)} optimal template bank is achieved by the $A_n^*$
  lattice, which is the best lattice-covering in dimensions $n\le 5$,
  and is close to the best covering known for dimensions $n \le 16$. 
  Generally this provides a \emph{substantially} more efficient
  covering than the simpler hyper-cubic lattice.
  We present an algorithm for generating lattice template banks for
  constant-coefficient metrics and we illustrate its implementation
  by generating $A_n^*$ template banks in $n=2,3,4$ dimensions.   
\end{abstract}



\section{Introduction}

The detection of gravitational waves (GWs) in the noisy data of detectors
ideally requires the knowledge of the signal waveform, in order
to coherently correlate the data with the expected signal by
\emph{matched filtering}.  Depending on the type of astrophysical
sources considered, however, one typically only knows a parametrized
family of possible waveforms (or approximations thereof).
The unknown parameters of these waveforms could be, for example, the
frequency and sky-position of spinning neutron stars, or the masses
and spins of inspiralling compact binary systems. 
Parameter spaces of such wide-parameter searches typically have
between one and four dimensions, depending on computational
constraints and the amount of astrophysical information available to
constrain the search space a-priori. In the case of GWs from general
binary systems, however, the number of dimensions of the parameter
space could be as large as 17.

Obviously one can only search a finite subset of points in this
parameter space, and this subset constitutes the ``template bank''.
The templates must \emph{cover} the parameter space, i.e., they must be
placed densely enough that no signal in this space can lose more than
a certain fraction of its power (called \emph{mismatch}) at the closest
template.     
However, coherently correlating the data with every template
is computationally expensive and increases the expected number of
statistical false-alarm candidates. 
An \emph{optimal} template bank therefore consists of the smallest
possible number of templates that still guarantees that the worst-case
mismatch does not exceed a given limit.  

It was realized early on that a geometric approach is very useful to
construct template banks, in particular the introduction of a
parameter-space \emph{metric} \cite{bala96:_gravit_binaries_metric,owen96:_search_templates}
based on the mismatch. 
This provides a natural measure of distance in parameter space and
allows one to ``correctly'' place templates, in the sense that the
maximal mismatch is not exceeded. 
Less attention, however, was devoted to the problem of \emph{optimally}
placing templates once the metric is known. Early works have sometimes
used a hyper-cubic template grid for illustrative purposes
\cite{owen96:_search_templates}, or the problem was incorrectly
referred to as a ``sphere packing problem''
\cite{1999PhRvD..60b2002O,brady98:_search_ligo_periodic}.  
We see in the following that constructing an optimal template bank is
an instance of the \emph{sphere covering} problem, which is somewhat
``dual'' to the sphere packing problem.  
The full solution to the sphere covering problem in Euclidean space 
is only known in $n=2$ dimensions, partial solutions
(restricted to lattices) are known in $n\le 5$ dimensions, while an
optimal solution for higher dimensions is unknown (cf.\ 
\cite{con99:_covering,schuerman06:_covering}).  
The main motivation of the present work is to develop a general method
for constructing efficient template banks in dimensions $n \lesssim 17$
by using the known results about Euclidean sphere covering.  

Previous related work on template banks includes studies to optimally
cover  \emph{non-flat} two-dimensional parameter spaces arising in
searches for GWs from inspiralling compact binary systems
\cite{2005CQGra..22.4285B,2006CQGra..23.5477B}. An interesting
algorithm to construct a hexagonal ($A_2^*$) template bank for
2D inspiral searches was described recently in \cite{2007arXiv0706.4437C}.
Various codes exist within the LIGO Scientific Collaboration
to generate hyper-cubic lattices (\texttt{LALCreateFlatMesh()} \cite{lalapps}), 
two-dimensional grids for non-constant metrics
(\texttt{LALCreateTwoDMesh()} \cite{lalapps}), and a tree-dimensional 
template bank based on the bcc-lattice
(\texttt{LALInspiralSpinBank()} \cite{lalapps}), which is being used
in a search for spinning binary inspirals on LIGO Data 
\cite{lscXX:_sbbh_S3_LIGO}. 

\section{Template-based searches and parameter-space metric}
\label{sec:templ-search-param}

A wide class of searches for GWs can be characterized as
\emph{template based}, in the sense that one searches for signals
belonging to a family of waveforms $s(t; \vparm)$, which depend on a
vector of parameters $\{\vparm\}^i = \parm^i$.
The strain $x(t)$ measured by a detector contains (usually dominating)
noise $n(t)$ in addition to possible weak GW signals
$s(t;\vparm_\sig)$, i.e., $x(t) = n(t) + s(t; \vparm_\sig)$.
One typically constructs a \emph{detection statistic}, $\F(\vparm;x)$ say,
namely a scalar characterizing the  probability of a
signal with parameters $\vparm$ being present in the data $x(t)$. 
Due to the random noise fluctuations $n(t)$ in the data, the detection
statistic is a random variable, and generally (assuming $\F$ is unbiased) 
its expectation value $\Fbar(\vparm;\vparm_\sig) \equiv E[\F(\vparm;x)]$ 
has a (local) maximum at the location of the signal $\vparm = \vparm_\sig$, i.e.,
\begin{equation}
  \label{eq:3}
  \left.\frac{\partial \Fbar(\vparm;\vparm_\sig)}{\partial \vparm}\right|_{\vparm=\vparm_\sig}
  = 0\,.
\end{equation}
Taylor-expanding the expected detection-statistic $\Fbar$ 
in small offsets $\Delta \vparm = \vparm - \vparm_\sig$
around the signal location $\vparm_\sig$ therefore reads as
\begin{equation}
  \label{eq:4}
  \Fbar(\vparm;\vparm_\sig) = \Fbar(\vparm_\sig; \vparm_\sig) 
  + \frac{1}{2}\left. \frac{\partial^2 \Fbar(\vparm;\vparm_\sig)}
    {\partial\parm^i\,\partial\parm^j}\right|_{\vparm_\sig}\,
  \! \Delta\parm^i \,\Delta\parm^j + \O(\Delta\parm^3)\,,
\end{equation}
where the matrix of second derivatives of $\Fbar$ is negative
definite. Here and in the
following we use automatic summation over repeated parameter indices
$i, j, \ldots$.
We can introduce a \emph{mismatch} $\mis$, which characterizes
the fractional loss in the expected value of the detection statistic,
$\Fbar$, at a parameter-space point $\vparm$, with respect to the
signal location $\vparm_\sig$, namely 
\begin{equation}
  \label{eq:5}
  \mis(\vparm;\vparm_\sig) \equiv 
  \frac{\Fbar(\vparm_\sig;\vparm_\sig) - \Fbar(\vparm;\vparm_\sig)}{\Fbar(\vparm_\sig;\vparm_\sig)}\,.
\end{equation}
Using the local expansion (\ref{eq:4}), we find
\begin{equation}
  \label{eq:6}
  \mis(\vparm;\vparm_\sig) = g_{i j}(\vparm_\sig)\,
  \Delta\parm^i \Delta \parm^j + \O(\Delta\parm^3)\,,
\end{equation}
where we defined the positive-definite \emph{metric tensor}
$g_{i j} \equiv - \frac{1}{2}\partial_i\partial_j \Fbar$, 
and $\partial_i \equiv \partial/\partial \parm^i$.
When searching a parameter space $\Parm(\parm^i,\,g_{ij})$, we need to 
compute the detection statistic $\F(x;\vparm_\xi)$ for a
discrete set of templates $\vparm_\xi\in\Parm$. Generally one can
distinguish two different approaches to this problem: one is a
random sampling of $\Parm$ using Markov-chain Monte-Carlo (MCMC)
algorithms (e.g.\ see \cite{2004PhRvD..70b2001C,2005PhRvD..72d3005C}), 
and the other consists of constructing a template bank 
$\Temp \equiv \{\vparm_\xi\} \subset\Parm$ that \emph{covers} the
whole of $\Parm$, in the sense that no point $\vparm \in \Parm$ exceeds a given
maximal mismatch $\mis_\max$ to its closest template $\vparm_\xi \in \Temp$, 
i.e.,  
\begin{equation}
  \label{eq:7}
  \underset{\vparm\in\Parm}{\max} \; \underset{\vparm_\xi\in\Temp}{\min}
  \; \mis(\vparm;\vparm_\xi) \le \mis_\max\,.
\end{equation}
Here we focus on the construction of \emph{optimal} template banks,
namely those satisfying (\ref{eq:7}) 
with the smallest possible number of templates $\vparm_\xi$.
In the local metric approximation (\ref{eq:6}), each template
$\vparm_\xi$ \emph{covers} a region $B_\xi$ of parameter space, namely
\begin{equation}
  \label{eq:8}
  B_\xi = \left\{ 
    \vparm \in \Parm : 
    g_{i j}(\vparm_\xi)\, \Delta\parm^i \,\Delta\parm^j \le \mis_\max\,,\quad
    \Delta\vparm \equiv \vparm - \vparm_\xi
  \right\}\,,
\end{equation}
which is a \emph{sphere} of radius $R = \sqrt{\mis_\max}$ 
in the metric space $\Parm(\parm^i,\,g_{ij})$.  
We can therefore reformulate the definition of an optimal template bank
as the set of (overlapping) spheres of covering radius $R$ which \emph{cover}
the whole of $\Parm$ in the sense of (\ref{eq:7}) with the smallest
number of spheres. 
This is known as the \emph{sphere covering problem}
\cite{con99:_covering}, not to be confused with the
somewhat dual \emph{sphere packing problem}, which seeks to pack the
largest number of non-overlapping ``hard'' spheres into a given volume. 
%

\section{The Euclidean sphere covering problem}
\label{sec:sphere-cover-probl}

In this section we summarize the current status of the sphere covering
problem as far as relevant for the construction of optimal template banks. 
There has been impressive progress in the study of the covering
problem in recent years, e.g.\ see \cite{con99:_covering} for a
general overview and \cite{schuerman06:_covering} for a more recent
update. Unfortunately, all of these studies are restricted to
Euclidean spaces $\Eucl^n$, while the metric parameter spaces
of GW searches are often curved.
In the following we will therefore make the assumption that 
$\Parm(\parm^i,\,g_{ij})$ can be treated as at least \emph{approximately} 
flat, or can be broken into smaller pieces that can be treated as
nearly flat. 
If the curvature of the metric is too strong, i.e., if the curvature radius
is comparable to the covering radius, it will be difficult to make use of
the Euclidean covering problem, and a different approach such as a
stochastic template bank or an MCMC sampling might be more effective.
We further assume that we have found a coordinate
system of $\Parm$ such that the metric components are (approximately)
constant, i.e., $g_{ij}(\vparm) \approx \const_{ij}$,
and for simplicity of notation we assume in this section (without
loss of generality) that we have chosen coordinates $x^i$ in which the
constant-coefficient metric is Cartesian, i.e.,  
$\Parm = \Eucl^n(x^i,\,\delta_{ij})$.

A covering can consist of any arrangement of covering spheres, but
currently all best coverings known are \emph{lattices}, and we
therefore restrict the following discussion to \emph{lattice coverings}.

\subsection{Basics on lattices}
\label{sec:basics-lattices}

An $n$-dimensional lattice $\Lat$ can be defined as a discrete set of
points $\dvec{\nu}_\xi$ (forming an additive group) generated by 
\begin{equation}
  \label{eq:10}
  \dvec{\nu}_\xi = \xi^i \,\dvec{l}_{(i)}\,,\quad\mbox{with}\quad \xi^i \in \Z\,,
\end{equation}
with summation over $i = 1, \ldots, n$, and 
where $\{\dvec{l}_{(i)}\}_{i=1}^n$ is a \emph{basis} of the lattice. Note
that it is sometimes convenient to express the $n$ basis vectors 
in a higher-dimensional Euclidean space, i.e., generally we can have 
$\dvec{l}_{(i)}\in \Eucl^m$  with $m \ge n$. When writing $\Eucl^n$ in
the following we refer to the subspace of $\Eucl^m$ containing the
$n$-dimensional lattice $\Lat$.
The $m \times n$ matrix ${M^a}_i \equiv l_{(i)}^a$ is called a \emph{generator matrix} 
of the lattice, with the columns of $M$ holding the $m$ components of the
$n$ lattice basis vectors, so we can also write the lattice $\Lat$ as
\begin{equation}
  \label{eq:40}
  \Lat = \left\{ \dvec{\nu}_\xi : \dvec{\nu}_\xi = M\,\dvec{\xi}\,,\;\;\dvec{\xi}\in \Z^n \right\}\,.
\end{equation}
The $n\times n$ matrix $A \equiv M^\tr\, M$ is called the \emph{Gram matrix}
(where $\tr$ denotes the transpose), which is symmetric and positive
definite, and  
$A_{ij} = \dvec{l}_{(i)}\cdot\dvec{l}_{(j)} = \delta_{ab} \,l_{(i)}^a \,l_{(j)}^b$, 
i.e., its coefficients are the mutual scalar products of lattice basis vectors.
\begin{figure}[htbp]
  \centering
  \includegraphics[width=0.65\textwidth]{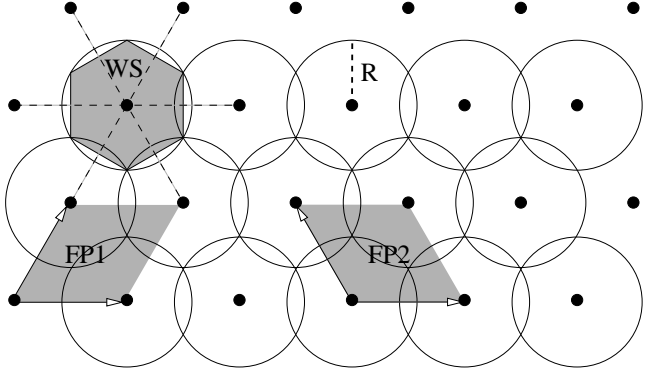}
  \caption{Hexagonal lattice ($A_2^*$) illustrating a 2-dimensional
    lattice covering. The shaded areas are different choices of
    fundamental regions for the lattice. FP1 and FP2 are 
    fundamental polytopes (\ref{eq:11}) associated with different
    choices of lattice basis, WS is the Wigner-Seitz cell
    (\ref{eq:14}), and $R$ is the covering radius.}
  \label{fig:An2}
\end{figure}
Each choice of lattice basis $\{\dvec{l}_{(i)}\}$ defines a
corresponding fundamental parallelotope ($\FP$), namely 
\begin{equation}
  \label{eq:11}
  \FP\left(\{\dvec{l}_{(i)}\}\right) \equiv \left\{ \dvec{x} \in \Eucl^n: \dvec{x} = \theta^i \,
    \dvec{l}_{(i)}\,,\;\; 0\le \theta^i < 1 \right\}\,,
\end{equation}
which is illustrated in \fref{fig:An2}. The $\FP$ is an example of a
\emph{fundamental region} for the lattice, i.e., a building block
containing exactly one lattice point, which fills the whole space
$\Eucl^n$ when repeated. There are many different choices of basis and
fundamental regions for the same lattice $\Lat$, but they all have the
same volume $\vol(\Lat)$, given by  
\begin{equation}
  \label{eq:13}
  \vol(\Lat) = \sqrt{\det A}\,,
\end{equation}
and in the case where $M$ is a square matrix we also have $\vol(\Lat) = \det M$.
One special choice of fundamental region is the \emph{nearest-neighbor region}, 
often referred to as \emph{Dirichlet-Voronoi} cell by mathematicians,
and more commonly known as  Wigner-Seitz cell or Brillouin zone by physicists,
which is defined as
\begin{equation}
  \label{eq:14}
  \WS(\Lat) \equiv \left\{ \dvec{x} \in \Eucl^n : \Vert \dvec{x}
    -\dvec{\nu}_0 \Vert \le \Vert \dvec{x} - \dvec{\nu}_\xi \Vert,\;\;
    \mbox{for all}\;\;\dvec{\nu}_\xi\in \Lat \right\}\,, 
\end{equation}
where $\Vert \dvec{x}\Vert = \sqrt{\dvec{x}\cdot\dvec{x}}\,$ is the
standard Euclidean norm in $\Eucl^n$. 
The vertices of the Wigner-Seitz cell are by construction local maxima
of the distance function of points in $\Eucl^n$ from the nearest grid
point. The maximum distance of any point in $\Eucl^n$ to the nearest
point of the lattice is called the \emph{covering radius} $R$,
which corresponds to the \emph{circumradius} of $\WS$, as
seen in \fref{fig:An2}.

Two lattices $\Lat_1$ and $\Lat_2$ with generator matrices $M_1$ and
$M_2$ are \emph{equivalent} if they can be transformed into one
another by a rotation, reflection and change of scale, namely if the
generator matrices satisfy 
\begin{equation}
  \label{eq:15}
  M_2 = c\, B\, M_1 \, U\,,
\end{equation}
where $c\in\R$ is a scale-factor, $U$ is integer-valued $\det U = \pm 1$,
which accounts for different choices of basis vectors, and $B$ is a real
orthogonal matrix, i.e., $B^\tr\,B = \unit$. The associated Gram
matrices are therefore related by  
\begin{equation}
  \label{eq:16}
  A_2 = c^2\, U^\tr\,A_1\,U\,,
\end{equation}
and the fundamental volumes (\ref{eq:13}) of the two lattices are
\begin{equation}
  \label{eq:19}
  \vol(\Lat_2) = c^n\, \vol(\Lat_1)\,.
\end{equation}
Let us consider as an example the 2-dimensional hexagonal lattice,
illustrated in \fref{fig:An2}. An obvious generator matrix is  
\begin{equation}
  \label{eq:17}
  M_1 = \left(\begin{array}{rr}
      1  & 1/2 \\
      0  & \sqrt{3}/2 \\
      \end{array} \right)\,,
\end{equation}
corresponding to $\FP1$ in \fref{fig:An2}.
However, sometimes it is more convenient to work with a generator
matrix of the form
\begin{equation}
  \label{eq:18}
  M_2 = \left(\begin{array}{rr}
      1  & 0 \\
      -1 & 1 \\
      0  & -1 \\
      \end{array}\right)\,,
\end{equation}
which has simpler coefficients, but uses a 3-dimensional representation
of the 2-dimensional lattice with all lattice points lying in the
plane $x+y+z=0$. 
One can verify that these two representations are equivalent in the
sense of (\ref{eq:15}), namely with 
\begin{equation}
  \label{eq:20}
  c = \sqrt{2}\,, \quad U = \left(\begin{array}{rr} 1 & 0\\ 0 & -1 \\ \end{array}\right)\,,\quad
  B = \left(\begin{array}{rr} 1/\sqrt{2} & - 1/\sqrt{6}\\ 
      -1/\sqrt{2} & -1/\sqrt{6} \\ 0 & \sqrt{2/3} \\ \end{array}\right)\,.
\end{equation}
Such a higher-dimensional representation of the generator matrix will
be useful later for the description of the n-dimensional $A_n^*$ lattice. 

\subsection{Known results on optimal sphere covering}
\label{sec:known-results-optim}

The efficiency of a sphere covering can be characterized by its
\emph{thickness} $\Theta$ 
(sometimes also referred to as the \emph{covering density}), which
measures the fractional amount of overlap between the covering
spheres, or equivalently the average number of spheres covering any point in $\Eucl^n$. 
This can be expressed as the ratio of the volume of one covering
sphere to the volume of the fundamental region of the lattice, i.e., 
\begin{equation}
  \label{eq:21}
  \Theta \equiv \frac{ V_n \, R^n}{\vol(\Lat)} \ge 1 \,,
\end{equation}
where $R$ is the covering radius and $V_n$ is the volume of the
unit-sphere in $n$ dimensions, namely $V_n = {\pi^{n/2}}/{\Gamma(n/2 +1)}$. 
We also use the \emph{normalized thickness} or \emph{center density}
$\theta$, defined as   
\begin{equation}
  \label{eq:22}
  \theta \equiv \frac{\Theta}{V_n}\,,  
\end{equation}
which corresponds to the number of centers (i.e., templates) per unit volume
in the case of $R=1$.  
Note that under a lattice transformation (\ref{eq:15}), the covering
radius $R$ obviously scales as $R_2 = c\,R_1$, and we therefore
see from (\ref{eq:19}) that the thickness (\ref{eq:21}) and
(\ref{eq:22}) is an invariant property of a lattice, i.e.,   
$\theta_2 = \theta_1$. 
The covering problem consists of finding the covering with the lowest
center density $\theta$.

Kershner showed in 1939 (see \cite{con99:_covering}) that in $n=2$
dimensions the most economical arrangement of circles covering the
plane is the hexagonal lattice, which is equivalent to an $A_2^*$ lattice.
In dimensions $n=3,4,5$ only the best \emph{lattice covering} is
known, and is given by $A_n^*$ in all three cases. 
In three dimensions, $A_3^*$ is also known as the 
\emph{body-centered-cubic} (bcc) lattice. Note that the best
\emph{packing} in $n=2$ is also achieved by the hexagonal lattice, but
for $n=3$ the face-centered cubic (fcc) lattice provides a denser
packing than bcc. 
In higher dimensions the best lattice coverings are currently still unknown,
but the best coverings known can be found in see table~2 of
\cite{schuerman06:_covering}, and \cite{schuerman07:_online}
provides for an up-to-date online version. As will become clearer in
the following, the $A_n^*$ lattice, while no longer the ``record
holder'' for most dimensions $5 < n \le 17$, is still close to the
best currently known covering in all cases. In the following we will
therefore mostly focus on the  $A_n^*$ covering.
The  $A_n^*$ lattice has a center density of 
\begin{equation}
  \label{eq:24}
  \theta(A_n^*) = \sqrt{n+1} \left\{ \frac{ n ( n + 2 ) }{12 (n + 1)}\right\}^{n/2}\,,
\end{equation}
while for the hyper-cubic grid $\Z^n$ the
Wigner-Seitz cell is a unit hypercube, so $\vol(\Z^n) = 1$, 
and the covering radius  $R = \sqrt{n}/2$ is half the length of the
diagonal. Therefore the center density (\ref{eq:22}) is found as 
$\theta(\Z^n) = 2^{-n}\, n^{n/2}$, 
which is dramatically worse than $A_n^*$ in higher dimensions, as can
be seen from the thickness ratio
\begin{equation}
  \label{eq:26}
  \kappa(n) \equiv \frac{\theta(\Z^n)}{\theta(A_n^*)} =
  \frac{3^{n/2}}{\sqrt{n+1}}\left( \frac{n + 1}{n+2}\right)^{n/2}
  \stackrel{n\rightarrow\infty}{\sim} \frac{3^{n/2}}{\sqrt{n e}}\,.
\end{equation}
\begin{table}[htbp]
  \caption{Thickness ratio $\kappa(n) = \theta(\Z^n) / A_n^*$,
    and $\gamma(n) = \theta(\mbox{best})/\theta(A_n^*)$
    in dimensions $n \le 17$.}   
  \label{tab:thickness} 
  \begin{indented}
    \item[]\begin{tabular}{@{} ccccccccc}
        \br
      $n$ 	& 2	& 3	& 4	& 5	& 6	& 7	& 8	&9	\\
      \mr
      $\kappa(n)$ & 1.3	&1.9	&2.8	&4.3	&6.8	&10.9	&17.7	&28.9	\\
      $\gamma(n)$ & 1.0	&1.0	&1.0	&1.0	&0.97	&0.95	&0.86   &0.97	 \\[0.2cm]
      \br
      $n$	&10	&11	&12	&13	&14	&15     &16	&17	\\
      \mr
      $\kappa(n)$ &47.4	&78.2	&130	&216	&359	&601	&1007	&1692	\\
      $\gamma(n)$ &0.98	&0.88	&0.99	&0.86	&0.82	&0.86	&1.0	&0.68\\
      \br
    \end{tabular}
  \end{indented}
\end{table}
There is a theoretical \emph{lower} limit on the thickness of any
covering, the Coxeter-Few-Rogers (CFR) bound $\tau_n$ (see
\cite{con99:_covering}), i.e., $\theta_n \ge \tau_n / V_n$, where
asymptotically $\tau_n \sim n/(e\sqrt{e})$ for $n \rightarrow \infty$. 
\begin{figure}[htbp]
  \centering
  \includegraphics[width=0.8\textwidth]{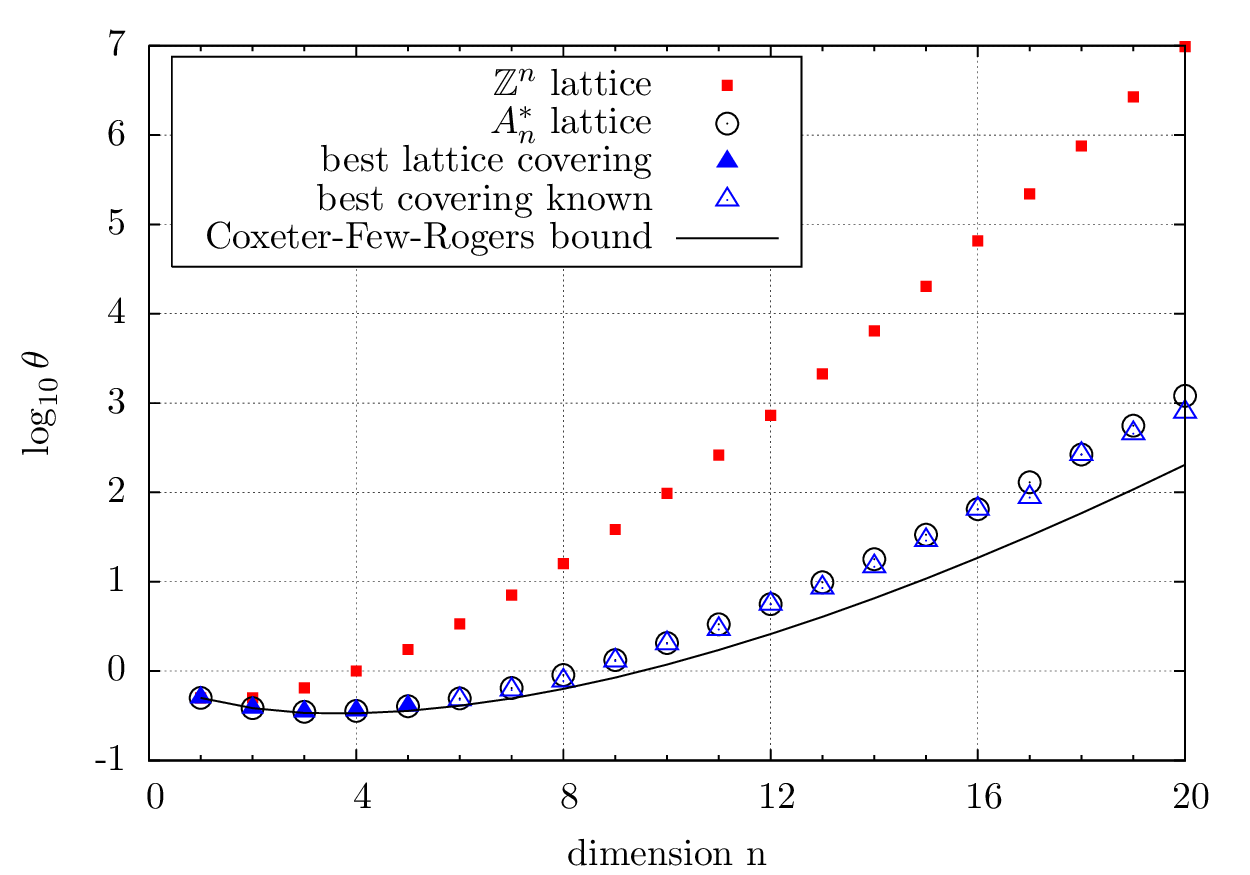}  
  \caption{Normalized covering thickness $\theta$  as function of dimension $n$,
    for the hyper-cubic lattice ($\Z^n$), the $A_n^*$ lattice,
    the theoretical lower bound (CFR), and the best lattice coverings 
    \emph{known}.}
  \label{fig:thickness}
\end{figure}
\Fref{fig:thickness} shows the normalized thickness $\theta$ as a
function of dimension $n$ for the $A_n^*$ and hyper-cubic $\Z^n$
lattices, as well as the CFR bound and the best covering known. 
In \tref{tab:thickness} we see that in dimensions $n>5$ where $A_n^*$
has been superseded as the best covering \cite{schuerman07:_online},
the relative improvement $\gamma(n) \equiv \theta(\mbox{best})/\theta(A_n^*)$ 
in thickness is typically quite small.
In particular, for $n\le16$ the improvement $\gamma(n)$ is typically less
than $18\%$, while the advantage $\kappa(n)$ of $A_n^*$ compared to
the hyper-cubic grid $\Z^n$ grows large very rapidly, as seen in
\tref{tab:thickness} and \fref{fig:thickness}. For practical
simplicity we therefore propose to use $A_n^*$ as the covering lattice of
choice.


\section{Lattice covering of template spaces}
\label{sec:impl-a_n-latt}

\subsection{Template counting}
\label{sec:template-counting}

The template spaces $\Parm(\parm^i,\,g_{ij})$
with constant-coefficient metrics $g_{ij}$ only differ
from the Cartesian case of the previous section
by a simple coordinate-transformation.
An infinitesimal parameter-space region $d^n\parm$ has a volume $d V$
measured by the metric, namely $d V = \sqrt{g}\, d^n\parm$,
where $g \equiv \det g_{ij}$. 
The volume $V$ of a finite region of parameter space is therefore 
\begin{equation}
  \label{eq:30}
  V = \int_{\Parm} d V = \sqrt{g}\,\int_\Parm d^n \parm\,,
\end{equation}
where we used the fact that $g_{ij}$ is a constant-coefficient metric.  
The number of templates $d N_p$ in $d V$ is given by
the inverse lattice volume, i.e., 
\begin{equation}
  \label{eq:31}
  d N_p = \frac{d V}{\vol(\Lat)}\,.
\end{equation}
Using the relation $R = \sqrt{\mis_\max}$ together with (\ref{eq:21}),
(\ref{eq:22}), we find  
\begin{equation}
  \label{eq:32}
  d N_p =  \theta\, \mis_\max^{-n/2}\, d V\,\quad\Longrightarrow\quad
  N_p = \theta \, \mis_\max^{-n/2}\, \sqrt{g}\, \int_\Parm d^n \parm\,,
\end{equation}
which generalizes template counting 
\cite{owen96:_search_templates,1999PhRvD..60b2002O,brady98:_search_ligo_periodic}     
to arbitrary lattices.

\subsection{Practical implementation of lattice covering}
\label{sec:latt-impl-templ}

In this section we present a practical algorithm for generating 
lattices covering of given maximal mismatch $\mis_\max$.
The approach described here works for any lattice generator $M$, but
in practice (cf.\ \sref{sec:known-results-optim}) we will be most
interested in the $A_n^*$ lattice. The generator for $A_n^*$ can be
expressed (cf.~\cite{con99:_covering}) as an $(n+1)\times n$ matrix\,,
\def\vsep{0.4em}
\begin{equation}
  \label{eq:27}
  {M^a}_j(A_n^*) = \left( \begin{array}{r r c r c}
      1   &   1  &  \ldots & 1  &  \frac{-n}{n+1} \\[\vsep]
     -1   &   0  &  \ldots & 0  &  \frac{1}{n+1}  \\[\vsep]
      0   &  -1  &  \ldots & 0  &  \frac{1}{n+1}  \\
 \vdots   &\vdots &  \vdots&\vdots& \vdots        \\
      0   &   0  &  \ldots & -1 &  \frac{1}{n+1}  \\[\vsep]
      0   &   0  &  \ldots & 0  &  \frac{1}{n+1}  \\
      \end{array}\right)\,,
\end{equation}
where the {columns} of $M$ hold the $n$ lattice basis
vectors $\dvec{l}_{(j)}$ expressed in $\Eucl^{n+1}$, i.e., ${M^a}_j = l_{(j)}^a$,
with index conventions $i, j = 1, \ldots, n$ and $a, b = 1, \ldots, n+1$.
The volume of the fundamental region and the covering radius for this
generator are 
\begin{equation}
  \label{eq:28}
  \vol(A_n^*) = \frac{1}{\sqrt{n+1}}\,,\quad\mbox{and}\quad 
  R(A_n^*) = \sqrt{\frac{n(n+2)}{12(n+1)}}\,,
\end{equation}
which yields the (normalized) thickness $\theta(A_n^*)$ given in (\ref{eq:24}).
In order to generate such a lattice in a parameter space
$\Parm(\parm^i,\,g_{ij})$, we need to express the generator ${M^a}_j$
in the $\parm^i$ coordinates, resulting in ${\Mt^i\,}_j$, say, such
that the lattice of templates $\dvec{\parm}_\xi$ is generated by  
\begin{equation}
  \label{eq:33}
  \parm^i_\xi = {\Mt^i\,}_j\, \xi^j\,,\quad\mbox{with}\quad \dvec{\xi} \in \Z^n\,.
\end{equation}
This coordinate transformation can be achieved in several steps:
\begin{enumerate}
\item Reduce the $(n+1)\times n$ matrix ${M^a}_j$ to a full rank
  generator, ${\Mh^i\,}_j$ say, by expressing the lattice basis
  vectors in a Euclidean basis spanning the $n$-dimensional subspace
  $\Eucl^n$ of the lattice: a simple Gram-Schmidt
  procedure with respect to the Cartesian metric $\delta_{ab}$ is used
  on the $\{l_{(j)}^a\}$ to generate an orthonormal basis $\{e_{(j)}^a\}$
  satisfying   
  \begin{equation}
    \label{eq:34}
    \delta_{a b} \, e_{(i)}^a \, e_{(j)}^b = \delta_{i j}\,.
  \end{equation}
The full-rank generator ${\Mh^i\,}_j$ is obtained from the components 
of the lattice vectors $\{l_{(i)}^a\}$ in this orthonormal basis, namely 
\begin{equation}
  \label{eq:35}
  {\Mh^i\,}_j  = \lh_{(j)}^{\,i} = l_{(j)}^a \, e_{(i)}^b \, \delta_{a b}
  = e_{(i) a}\, {M^a\,}_j \,.
\end{equation}

\item Translate the full-rank generator ${\Mh^i\,}_j$ from Cartesian
  coordinates into the coordinate system $\parm^i$ with metric $g_{ij}$.
  For this we use another Gram-Schmidt orthonormalization with respect to
  the metric $g_{ij}$, with the lattice vectors $\{\lh_{(i)}^{\,j}\}$ as
  input to find an orthonormal basis $\{d_{(i)}^j\}$ satisfying  
  \begin{equation}
    \label{eq:36}
    g_{ij} \, d^i_{(l)} \, d^{j}_{(k)} = \delta_{l k}\,.
  \end{equation}
  This representation of an orthonormal basis in coordinates $\parm^i$
  allows us to express the lattice vectors in these coordinates as 
  \begin{equation}
    \label{eq:37}
    \lt_{(j)}^i = \lh_{(j)}^{\,k} \, d_{(k)}^i = d_{(k)}^i\, {\Mh^k\,}_j\,.
  \end{equation}

\item Scale the generator to the desired covering radius 
  $R = \sqrt{\mis_\max}$, and with (\ref{eq:28}) we find
  \begin{equation}
    \label{eq:38}
    {\Mt^i\,}_j = \sqrt{\mis_\max}\, \sqrt\frac{12(n+1)}{n(n+2)} \,\, \lt_{(j)}^i\,,
  \end{equation}
  which is a generator (\ref{eq:33}) for an $A_n^*$
  template lattice with maximal mismatch $\mis_\max$. 
\end{enumerate}
This algorithm has been implemented in
\texttt{XLALFindCoveringGenerator()} in LAL \cite{lalapps}, and some
tests of this code are presented in the next section.

\subsection{Tests of the implementation}
\label{sec:test-implementation}

\begin{figure}[htbp]
  \centering
  \mbox{
    \includegraphics[height=5.8cm]{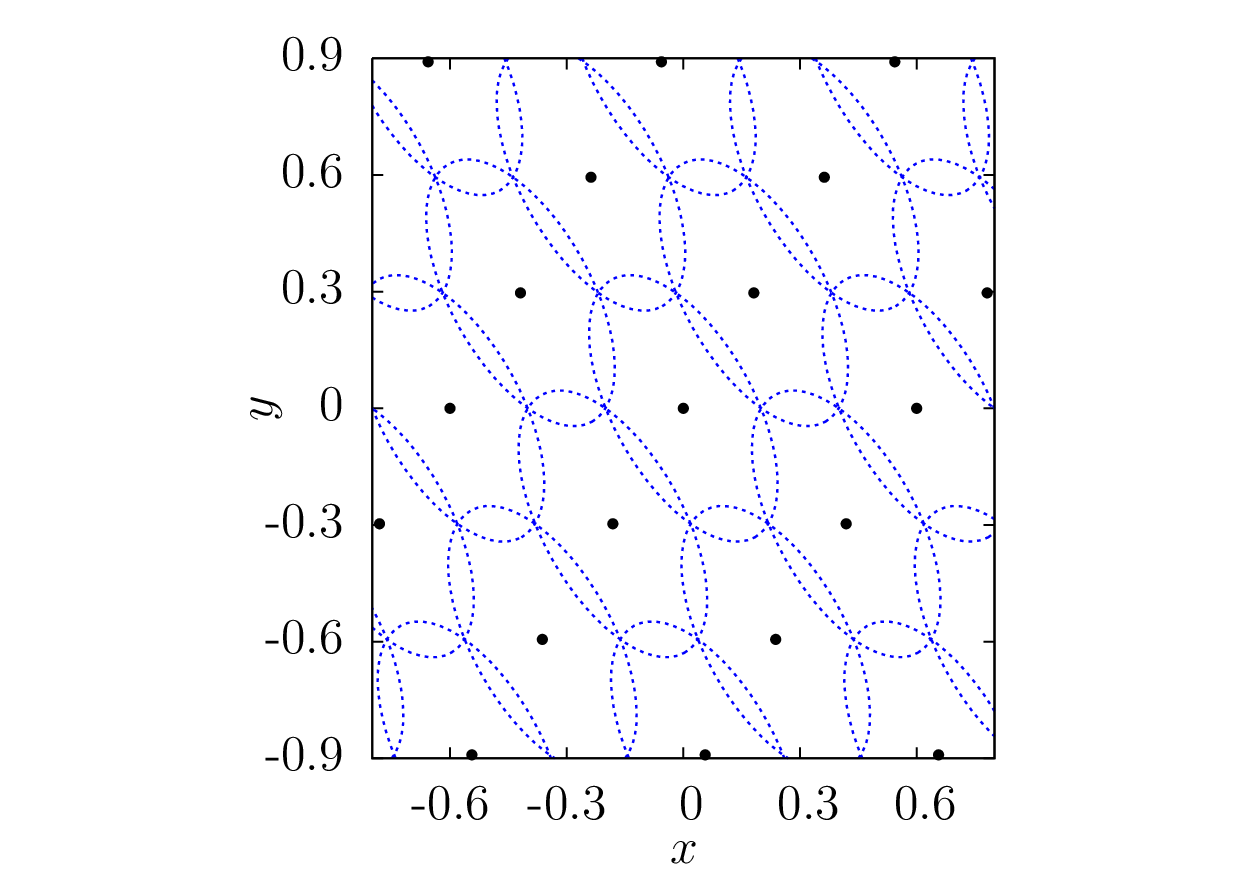}
    \hspace*{0.1cm}
    \includegraphics[height=5.8cm]{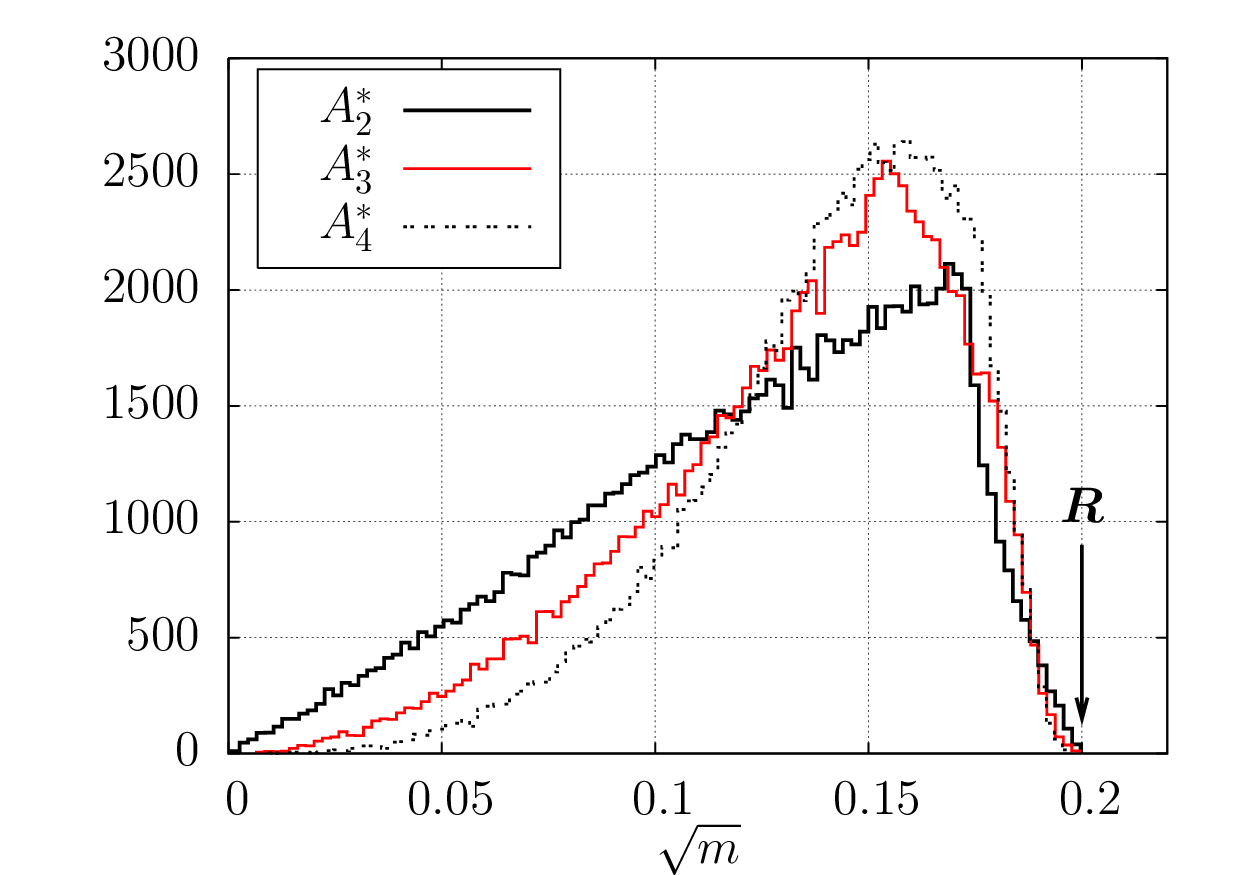}
  }
  \caption{\emph{Left panel:} Hexagonal ($A_2^*$) lattice covering
    in coordinates $\{x,y\}$ with metric $g_{ij} = [ 1,\,0.4;\,0.4,\,0.5 ]$.  
    \emph{Right panel:} Histogram of measured distances $\sqrt\mis$ in
    a Monte-Carlo sampling of 100,000 points from an $A_n^*$ covering
    in $n=2,3,4$ dimensions, using non-Cartesian metrics $g_{ij}$. 
    The nominal covering radius in all three cases was $R = \sqrt{\mis_\max} = 0.2$. 
  }
  \label{fig:Ans2D}
\end{figure}

In order to illustrate and test the implementation of this algorithm,
we generate an $A_n^*$ lattice in dimensions $n=2, 3, 4$,   
respectively, with a maximal mismatch of $\mis_\max = 0.04$, i.e., a
covering radius of $R = 0.2$.  
For generality we use a non-Cartesian metric $g_{ij} \not= \delta_{ij}$, 
as illustrated in the left panel of \fref{fig:Ans2D}. 
We picked 100,000 points $\vparm\in\Parm(\parm^i,\,g_{ij})$ at
random and computed their mismatch $\mis$ (using the metric) to the
nearest template $\vparm_\xi$, which is a way of \emph{measuring} the
maximal mismatch of a template bank. The distribution of measured
mismatch-distances $\sqrt{\mis}$ is plotted in the right-hand panel of
\fref{fig:Ans2D}, and we see that the mismatches are bounded by
$\sqrt{\mis_\max}=0.2$, satisfying (\ref{eq:7}). 
We can also measure the (normalized) thickness $\theta$ of the
template bank, namely from the number of templates $N_p$ in the
covered parameter space $\Delta\vparm^n$, we find using (\ref{eq:32}):
\begin{equation}
  \label{eq:39}
  \theta = \frac{R^n}{\sqrt{g}}\, \frac{N_p}{\Delta \vparm^n}\,.
\end{equation}
These measured values of the thickness are found to agree to within
$0.2\%$ with the theoretical values (\ref{eq:24}) in all three cases $n=2,3,4$.
The generated template banks in this example have $N_p \sim \O(10^4)$
templates, and the error can most likely be attributed to boundary effects.   

\section{Discussion}
\label{sec:discussion}

Possible applications of this algorithm for GW searches can be found in
template-based searches, such as for inspiralling compact binary systems 
and for ``continuous waves'', which in ground-based detectors refers
mostly to signals from spinning neutron stars, and in the case of LISA
includes white dwarf binaries, supermassive black hole binaries and
extreme-mass ratio inspirals. The benefit of using this approach
depends sensitively on the number of parameter-space dimensions, but
can be estimated from \tref{tab:thickness} at least in comparison to
hypercubic grids.

However, the applicability of the lattice covering algorithm presented here is
restricted to \emph{explicitly flat} parameter spaces, which limits
its usefulness to cases where we can find a coordinate system in which
the parameter-space metric is (at least) \emph{approximately} constant.
The orbital metric approximation \cite{prix06:_searc} for
continuous GWs can be shown to be flat (work in progress), and would
therefore be a natural case where this lattice covering could be 
used to greatest effect.
One difficulty in this case, however, stems from that fact that the
corresponding metric is found to be highly ill-conditioned, which
results in the lattice-construction algorithm to fail due to numerical
problems.    
One therefore needs to \emph{analytically} ``factor out'' this
near-degeneracy of the metric before this lattice-covering procedure
can be safely applied. 
More work is also required to deal with non-trivial parameter-space
boundaries, which complicates the $n$-dimensional filling algorithm.  

\section*{Acknowledgments}
I am grateful for discussions with Bruce Allen, Andrzej Kr{\'o}lak and
Curt Cutler. I further thank Achill Sch\"urmann for reading the
manuscript and providing helpful comments. 
This work was supported by the Max-Planck-Society (\aei).
This paper has been assigned LIGO Document Number \dcc.

\section*{References}
\bibliography{biblio}

\end{document}